\documentstyle{article}
\begin{document}
\centerline{\Large \bf Paradoxes in the Boltzmann kinetic theory}
\vskip30pt
\centerline{C. Y. Chen}
\centerline{Department of Physics, Beijing University of Aeronautics}
\centerline{ and
Astronautics, Beijing, 100083, P.R.China}
\centerline{ Email: cychen@buaa.edu.cn}

\vskip 0.9cm
\begin{abstract} Paradoxes in the Boltzmann kinetic theory are presented. Firstly, it is pointed out that the usual notion concerning the perfect continuity of distribution function is not generally valid; in many important situations using certain types of discontinuous distribution functions is an absolute must. Secondly, it is revealed that there is no time reversibility in terms of beam-to-beam collisions and, in connection with this, there are intrinsic difficulties in formulating the net change of  molecular density due to collisions, either in the three-dimensional velocity space or in the six-dimensional phase space. With help of simple examples, the paradoxes manifest themselves clearly.
\end{abstract}

\vskip 1.2cm
\section{Introduction}
When the Boltzmann equation, thought of as the first equation of kinetic theory, came out, many philosophical doubts arose\cite{pathria}, of which most were related to the fact that while Newton's equations themselves were time-reversible the Boltzmann equation, as a consequence of Newton's equations, was time-irreversible. As time went by, in particular after the formally rigorous BBGKY theory was formulated in the middle of the last centrary[2$-$8], concerns of the time-reversal paradox gradually faded out, and it was believed that the ultimate understanding of the related issues had been completely established (at least in the regime of classical physics).

Nevertheless, later developments of mathematics and physics have suggested that things should be more complicated than they appear to be. As one of the developments, the studies of fractals\cite{man1,man2} illustrate that complex structures on very small scales resembling those on large scales exist quite generally. Impact of the fractal studies on our customary thought can be various and enormous: some of us may be led to think about the possibility that nature gets certain mechanisms which allow macroscopic structures to shape and form microscopic structures, and others may be tempted to construct new formalisms describing dynamics of complex structures over infinitely small ranges.  

As another of the developments, it has been revealed that the standard treatment of collisions involves essential problems and the collective behavior of molecular collisions should be reinvestigated\cite{chen0}.

In this paper, we wish to present a rather comprehensive description of paradoxical aspects of the Boltzmann equation. Firstly, it is shown that many mechanisms in a real gas are capable of producing abnormal molecules not associated with continuous distribution functions; as long as macroscopic nonequilibrium exists somewhere, sharp, even discontinuous, microscopic structures will result elsewhere and certain types of discontinuous distribution functions need to be employed.
Secondly, it is illustrated that formulating the number of molecules entering and leaving a volume element is a very tricky task to accomplish. Much contrary to the customary thought, there is no time reversibility that we can utilize. Even worse, if we want to know how many molecules enter and leave an infinitesimal phase volume element, we shall meet with conflicting differential processes. All these paradoxes point to one thing: it is not possible to have a formalism that is mainly based on differential-type analyses and can still describe kinetic dynamics in the phase space.

We shall here present only paradoxes that can be investigated from simple and mathematical analyzes, in view of
that a large number of philosophical issues, less mathematical but not less important, have been debated rather
intensively.

\section{A brief survey of the Boltzmann equation}
There are many descriptions of the Boltzmann equation in the literature[12-14]. For purposes of this paper, we shall recall the Boltzmann equation and its derivation in a brief and illustrative way.

The core of the Boltzmann equation, which also serves as the basics of other kinetic equations, says that if collisions of molecules of a gas can be neglected in a certain region, the distribution function there satisfies the collisionless Boltzmann equation
\begin{equation} \label{collisionless} \frac{\partial f}{\partial t} +
{\bf v} \cdot \frac{\partial f}{\partial {\bf r}}+\frac{{\bf F}}m \cdot \frac{\partial f}
{\partial {\bf v}} =0,
\end{equation}
which is sometimes called the continuity equation in the phase space. The proof of this continuity equation is obtained by considering a six-dimensional phase volume element (box-shaped) and formulating the total flux through all the facets of the volume element. According to textbooks, this equation is also equivalent to the path invariance of distribution function
\begin{equation} \label{invariance} \left.\frac{df}{dt}\right|_{{\bf r}(t),{\bf v}(t)}=0
\quad {\rm or}\quad f|_{{\bf r}(t),{\bf v}(t)}={\rm Constant},
\end{equation}
where the subindexes ${\bf r}(t)$ and ${\bf v}(t)$ indicate that that the expression is valid along a molecule's trajectory.
Conceptually speaking, equation (\ref{invariance}) can be interpreted as saying that such a gas is `incompressible' in the phase space.
The direct proof of the path invariance expressed by (\ref{invariance}) is usually associated with a Jacobian formulation. The proofs of (\ref{collisionless}) and (\ref{invariance}) in textbooks appear to be stringent except that there is one tacit assumption: the distribution function is perfectly continuous (differentiable, to be more exact) in terms of the six position and velocity components. Behind this continuity assumption, there is a more fundamental concept called molecular chaos\cite{cer}. It is explicitly or implicitly assumed that with help of molecular chaos the perfect continuity of all interested distribution functions is virtually ensured.

To incorporate collisional effects into the formalism, collisions between two molecular beams, with velocities ${\bf v}_1$ and ${\bf v}_2$ respectively, are first considered.
The number of type-1 molecules (with the velocity ${\bf v}_1$) leaving $d{\bf r}d{\bf v}_1$ during a time interval $dt$ is then found to be, according to the standard approach,
\begin{equation}\label{leave}
dtd{\bf r} d{\bf v}_1 \int_{{\bf v}_2,{\bf v}_1^\prime, {\bf v}_2^\prime}
u f({\bf v}_1)f({\bf v}_2) \sigma({{\bf v}_1, {\bf v}_2\rightarrow \bf v}_1^\prime, {\bf v}_2^\prime )
 d{\bf v}_2 d{\bf v}_1^\prime d{\bf v}_2^\prime,
\end{equation}
where ${\bf v}_1^\prime$ and ${\bf v}_2^\prime$ are the final velocities of type-1 and type-2 molecules respectively, $u=|{\bf v}_2^\prime-{\bf v}_1^\prime|=|{\bf v}_2-{\bf v}_1|$ and $\sigma( {\bf v}_1, {\bf v}_2\rightarrow  {\bf v}_1^\prime, {\bf v}_2^\prime)$ is the cross section in the laboratory frame; likewise, the number of the molecules entering
$d{\bf r}d{\bf v}_1$ during $dt$ to be
\begin{equation}\label{enter}
dtd{\bf r} d{\bf v}_1 \int_{{\bf v}_2,{\bf v}_1^\prime, {\bf v}_2^\prime}
u f({\bf v}_1^\prime) f({\bf v}_2^\prime) \sigma({\bf v}_1^\prime, {\bf v}_2^\prime\rightarrow
{\bf v}_1, {\bf v}_2)
 d{\bf v}_2 d{\bf v}_1^\prime d{\bf v}_2^\prime.
\end{equation}
In view of the conservation of molecular number, the standard approach assumes that
\begin{equation} \label{twosec}  \int_{\Omega}
\sigma(\Omega)d\Omega =\int_{{\bf v}_1^\prime}\int_{{\bf
v}_2^\prime} \sigma({\bf v}_1,{\bf v}_2\rightarrow {\bf
v}_1^\prime,{\bf v}_2^\prime)
 d{\bf v}_1^\prime d{\bf v}_2^\prime, \end{equation}
where $\sigma(\Omega)$ is the cross section defined in the center-of-mass frame.
The two cross sections in (\ref{leave}) and (\ref{enter}) are considered to be equal to each other because of the time-reversibility of collision.
On this understanding, the net change of the molecular number per unit phase volume and per unit time can be expressed by
\begin{equation}\label{change}  \int_{{\bf v}_2,\Omega}
u[f({\bf v}_1^\prime)f({\bf v}_2^\prime)-f({\bf v}_1)f({\bf v}_2)]\sigma
(\Omega)d\Omega d{\bf v}_2. \end{equation}
By regarding (\ref{change}) as a correction term of the collisionless
Boltzmann equation, the standard Boltzmann equation is obtained as
\begin{equation}\label{bl} \frac{\partial f}{\partial t}+{\bf v}_1\cdot
\frac{\partial f}{\partial {\bf r}} +\frac {{\bf F}}m
\cdot\frac{\partial f}{\partial {\bf v}_1}=\int_{{\bf v}_2,\Omega}
[f({\bf v}_1^\prime)f({\bf v}_2^\prime)-f({\bf v}_1)f({\bf
v}_2)]u\sigma (\Omega)d\Omega d{\bf v}_2. \end{equation}

The Boltzmann equation (\ref{bl}) is usually recognized as an independent
and complete dynamical description of a gas consisting of hard-sphere molecules.
The terms `independent' and `complete' mean that the solution for the equation can be constructed without getting help from other physical laws, such as Newton's equations for individual molecules. It is widely
believed that if sufficiently powerful computers were available now,
we would be able to solve the equation by applying certain types of finite
difference schemes (under certain boundary conditions and initial conditions, of course).

Although the Boltzmann equation and its derivations in textbooks seem quite stringent and have been accepted unanimously, many paradoxes related to it can be raised. Ironically enough, even a simple comparison between the left and right sides provides enlightening things to ponder. On the left side, molecules entering and leaving a six-dimensional phase volume element $d{\bf r} d{\bf v}_1$ have been considered and there is a good symmetry between the position vector ${\bf r}$ and the velocity vector ${\bf v}_1$ in terms of differential operations. On the right side, there is no symmetry between ${\bf r}$ and ${\bf v}_1$, all the integral operations are performed in the velocity space and the position vector ${\bf r}$ serves only as an inactive parameter. That is to say, although the derivation of the collisional operator is formally carried out in the phase space, only the velocity subspace has been in one's mind. If a forum were held in this community to discuss which space should be used in terms of formulating the collisional operator, a 50-to-50 situation would be there: about a half of us may prefer the phase space in view of that the left side of the Boltzmann equation is indeed written down in the phase space, while the rest may choose the velocity space in view of that collision issues have, almost always, been investigated in the velocity space only. In Sect.~4 of this paper, we shall return to this 
subject. Much to our surprise, it will be unveiled that, unless some of our basic concepts are changed, we shall have intrinsic difficulties in formulating collisional effects.

\section{The continuity aspect of distribution function}
Here, we wish to show that the perfect continuity of distribution function is just an assumption; in many important situations, using certain types of discontinuous distribution functions is absolutely a must. Throughout this section, in order to make the discussion less cumbersome, molecule-to-molecule collisions are considered negligible, at least in terms of the zeroth-order approximation.

We first establish the connection between the perfect continuity of distribution function and the path invariance expressed by (\ref{invariance}).
Though the six-dimensional phase space is somewhat abstract, we can visualize the path invariance in it quite easily.  Fig.~1 is drawn to show the situation in the two-dimensional $x-v_x$ phase space; but things taking place in it can schematically manifest what happens in the full six-dimensional phase space. Suppose that the molecules inside the rectangle in Fig.~1a distribute rather uniformly at $t=0$; these molecules will, due to molecular motion, distribute inside the parallelogram in Fig.~1b rather uniformly at $t=T$. By excluding all external forces (to make our understanding as simple as possible), we readily see that the area of the rectangle is equal to that of the parallelogram, which means that the molecular density inside the moving volume element keeps invariant. Based on the picture outlined above, we can use a sufficiently small phase volume element and let it move along a molecule's path, the average density of molecules in it will not change, irrespective of the shape and size of the employed volume element.

\hspace{-0.4cm}
\setlength{\unitlength}{0.022in}
\begin{picture}(100,90)

\multiput(15,19)(90,0){2}{\vector(1,0){80}}
\multiput(20,15)(90,0){2}{\vector(0,1){60}}

\put(45,40){\framebox(20,20){}}
\multiput(135,40)(20,20){2}{\line(1,0){20}}
\multiput(135,40)(20,0){2}{\line(1,1){20}}

\multiput(48,42.5)(2,0){8}{\circle*{1.0}}
\multiput(47,45.5)(2,0){9}{\circle*{1.0}}
\multiput(48,48.5)(2,0){8}{\circle*{1.0}}
\multiput(47,51.5)(2,0){9}{\circle*{1.0}}
\multiput(48,54.5)(2,0){8}{\circle*{1.0}}
\multiput(47,57.5)(2,0){9}{\circle*{1.0}}

\multiput(139.5,42.5)(2,0){8}{\circle*{1.0}}
\multiput(143,45.5)(2,0){9}{\circle*{1.0}}
\multiput(145.5,48.5)(2,0){8}{\circle*{1.0}}
\multiput(149,51.5)(2,0){9}{\circle*{1.0}}
\multiput(151.5,54.5)(2,0){8}{\circle*{1.0}}
\multiput(155,57.5)(2,0){9}{\circle*{1.0}}

\put(50,5){\makebox(20,8)[l]{\bf (a)}}
\put(140,5){\makebox(20,8)[l]{\bf (b)}}

\multiput(89,15)(90,0){2}{\makebox(20,8)[c]{$x$}}
\multiput(10,77)(90,0){2}{\makebox(20,8)[c]{$v_x$}}

\put(46,32){\makebox(20,8)[c]{\small $\Delta x$}}
\put(27,47){\makebox(20,8)[c]{\small $\Delta v_x$}}
\end{picture}
\vspace{-0.4cm}
\begin{center}
\begin{minipage}{10cm}
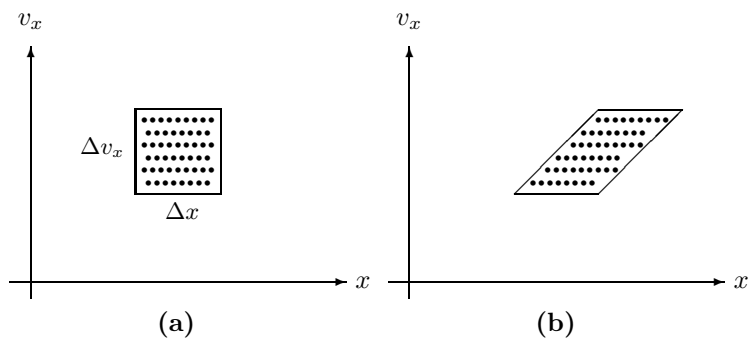
{Figure~1: Schematic of molecular motion in the phase space.
(a) A number of molecules distribute inside the rectangle at $t=0$. (b) They will distribute inside a paralellogram at $t=T$.}
\end{minipage}
\end{center}

\hspace{-0.4cm} 
\setlength{\unitlength}{0.022in}
\begin{picture}(100,90)
\multiput(15,19)(90,0){2}{\vector(1,0){80}} \multiput(20,15)(90,0){2}{\vector(0,1){60}}
\put(45,40){\framebox(20,20){}} \multiput(45,40)(1.3,1.7){4}{\circle*{1.2}} \put(50,46.4){\circle*{1.2}}
\multiput(51.4,48)(1.53,1.53){5}{\circle*{1.2}} \multiput(59.2,55.8)(1.7,1.3){4}{\circle*{1.2}}
\multiput(135,40)(20,20){2}{\line(1,0){20}} \multiput(135,40)(20,0){2}{\line(1,1){20}}
\multiput(135,40)(3,1.7){4}{\circle*{1.2}} \multiput(147,47.2)(3,1.5){6}{\circle*{1.2}}
\multiput(165,56.2)(3.1,1.2){4}{\circle*{1.2}} \put(50,5){\makebox(20,8)[l]{\bf (a)}}
\put(140,5){\makebox(20,8)[l]{\bf (b)}} \multiput(89,15)(90,0){2}{\makebox(20,8)[c]{$x$}}
\multiput(10,77)(90,0){2}{\makebox(20,8)[c]{$v_x$}} \put(46,32){\makebox(20,8)[c]{\small $\Delta x$}}
\put(27,47){\makebox(20,8)[c]{\small $\Delta v_x$}}
\end{picture}

\begin{center}
\begin{minipage}{10cm}
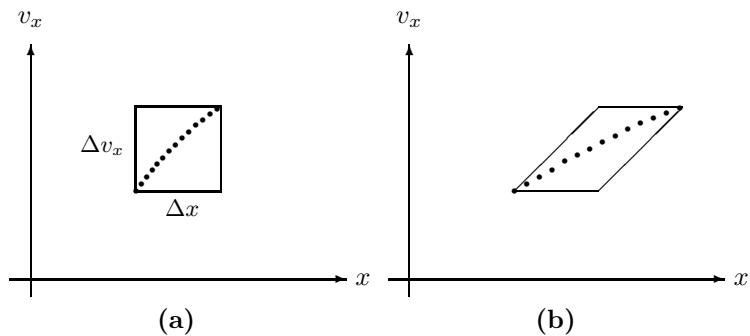
{Figure~2: For confined molecules, the situation can be much different from that in Fig.~1. (a) All molecules
distribute along a line at $t=0$. (b) Separations between the molecules become larger at $t=T$.}
\end{minipage}\end{center}

However, if the molecules are confined to a certain region in the phase space, the conclusion given in the last
paragraph will be valid no longer. For this to be seen, investigate Fig.~2a, in which all the molecules distribute
along one diagonal of the rectangle at $t=0$, and they will, at $t=T$, move to new places and distribute along the
diagonal of the parallelogram, as shown in Fig.~2b. If we have a phase volume element of a fixed size and shape
and let it move with the molecules, we find that the molecular density in it, defined as the molecular number
divided by the phase volume, changes with time. Furthermore, it can be seen that if the shape and size
of the employed volume element are allowed to change, we shall get different molecular densities, varying
drastically from zero to infinity.

A crucial question arises. Can molecules in a real gas be so confined? To find out the answer, consider a beam of molecules striking a surface $S$ and `reflected' from it; or consider a surface $S$ moving toward a group of stationary molecules and driving them to move. As shown in Fig.~3a, if $S$ is convex and relatively smooth, the molecules will, after colliding with the surface, spread out in space with definite velocities. (Even if the surface is not truly smooth, the picture shown will not change significantly.) By counting the molecules in a moving spatial volume element $\Delta {\bf r}$ and in a velocity range between ${\bf v}$ and ${\bf v}+\Delta {\bf v}$, we find that the molecular density within $\Delta {\bf r}\Delta {\bf v}$ decreases. This means that these molecules are diverging in the phase space or, in terms of the distribution function $f$ describing them, we get an expression as
\begin{equation} \label{co1} \left.\frac{df}{dt}\right|_{{\bf r}(t),{\bf v}(t)} <0
\quad {\rm or }\quad \frac{\partial f}{\partial t} +
{\bf v} \cdot \frac{\partial f}{\partial {\bf r}}+\frac{{\bf F}}m \cdot \frac{\partial f}
{\partial {\bf v}} <0.
\end{equation}
For the situation shown in Fig.~3b, we may have, in a certain region along molecular paths,
\begin{equation} \label{co2} \left.\frac{df}{dt}\right|_{{\bf r}(t),{\bf v}(t)} >0.
\end{equation}
Equations (\ref{co1})-(\ref{co2}) and Fig.~3 tell us that diverging and converging molecules associated with discontinuous distributions can be constantly produced by collisions between molecules and solid boundaries and these discontinuous distributions will keep discontinuous along the molecular paths.

\hspace{0.45cm}
\setlength{\unitlength}{0.018in}
\begin{picture}(400,105)
\put(25.000, 55.000){\circle*{1}}
\put(24.986, 55.600){\circle*{1.0}}
\put(24.986, 54.400){\circle*{1.0}}
\put(24.971, 56.200){\circle*{1.0}}
\put(24.971, 53.800){\circle*{1.0}}
\put(24.957, 56.799){\circle*{1.0}}
\put(24.957, 53.201){\circle*{1.0}}
\put(24.942, 57.399){\circle*{1.0}}
\put(24.942, 52.601){\circle*{1.0}}
\put(24.928, 57.999){\circle*{1.0}}
\put(24.928, 52.001){\circle*{1.0}}
\put(24.914, 58.599){\circle*{1.0}}
\put(24.914, 51.401){\circle*{1.0}}
\put(24.873, 59.198){\circle*{1.0}}
\put(24.873, 50.802){\circle*{1.0}}
\put(24.832, 59.796){\circle*{1.0}}
\put(24.832, 50.204){\circle*{1.0}}
\put(24.792, 60.395){\circle*{1.0}}
\put(24.792, 49.605){\circle*{1.0}}
\put(24.751, 60.993){\circle*{1.0}}
\put(24.751, 49.007){\circle*{1.0}}
\put(24.710, 61.592){\circle*{1.0}}
\put(24.710, 48.408){\circle*{1.0}}
\put(24.670, 62.191){\circle*{1.0}}
\put(24.670, 47.809){\circle*{1.0}}
\put(24.629, 62.789){\circle*{1.0}}
\put(24.629, 47.211){\circle*{1.0}}
\put(24.558, 63.385){\circle*{1.0}}
\put(24.558, 46.615){\circle*{1.0}}
\put(24.487, 63.981){\circle*{1.0}}
\put(24.487, 46.019){\circle*{1.0}}
\put(24.416, 64.577){\circle*{1.0}}
\put(24.416, 45.423){\circle*{1.0}}
\put(24.344, 65.172){\circle*{1.0}}
\put(24.344, 44.828){\circle*{1.0}}
\put(24.273, 65.768){\circle*{1.0}}
\put(24.273, 44.232){\circle*{1.0}}
\put(24.202, 66.364){\circle*{1.0}}
\put(24.202, 43.636){\circle*{1.0}}
\put(24.104, 66.956){\circle*{1.0}}
\put(24.104, 43.044){\circle*{1.0}}
\put(24.006, 67.548){\circle*{1.0}}
\put(24.006, 42.452){\circle*{1.0}}
\put(23.907, 68.140){\circle*{1.0}}
\put(23.907, 41.860){\circle*{1.0}}
\put(23.809, 68.731){\circle*{1.0}}
\put(23.809, 41.269){\circle*{1.0}}
\put(23.711, 69.323){\circle*{1.0}}
\put(23.711, 40.677){\circle*{1.0}}
\put(23.612, 69.915){\circle*{1.0}}
\put(23.612, 40.085){\circle*{1.0}}
\put(23.514, 70.507){\circle*{1.0}}
\put(23.514, 39.493){\circle*{1.0}}
\put(23.384, 71.093){\circle*{1.0}}
\put(23.384, 38.907){\circle*{1.0}}
\put(23.253, 71.678){\circle*{1.0}}
\put(23.253, 38.322){\circle*{1.0}}
\put(23.123, 72.264){\circle*{1.0}}
\put(23.123, 37.736){\circle*{1.0}}
\put(22.993, 72.850){\circle*{1.0}}
\put(22.993, 37.150){\circle*{1.0}}
\put(22.862, 73.436){\circle*{1.0}}
\put(22.862, 36.564){\circle*{1.0}}
\put(22.732, 74.021){\circle*{1.0}}
\put(22.732, 35.979){\circle*{1.0}}
\put(22.572, 74.600){\circle*{1.0}}
\put(22.572, 35.400){\circle*{1.0}}
\put(22.413, 75.178){\circle*{1.0}}
\put(22.413, 34.822){\circle*{1.0}}
\put(22.253, 75.756){\circle*{1.0}}
\put(22.253, 34.244){\circle*{1.0}}
\put(22.094, 76.335){\circle*{1.0}}
\put(22.094, 33.665){\circle*{1.0}}
\put(21.934, 76.913){\circle*{1.0}}
\put(21.934, 33.087){\circle*{1.0}}
\put(21.775, 77.492){\circle*{1.0}}
\put(21.775, 32.508){\circle*{1.0}}
\put(21.615, 78.070){\circle*{1.0}}
\put(21.615, 31.930){\circle*{1.0}}
\put(21.420, 78.637){\circle*{1.0}}
\put(21.420, 31.363){\circle*{1.0}}
\put(21.226, 79.205){\circle*{1.0}}
\put(21.226, 30.795){\circle*{1.0}}
\put(21.031, 79.773){\circle*{1.0}}
\put(21.031, 30.227){\circle*{1.0}}
\put(20.836, 80.340){\circle*{1.0}}
\put(20.836, 29.660){\circle*{1.0}}
\put(20.641, 80.908){\circle*{1.0}}
\put(20.641, 29.092){\circle*{1.0}}
\put(20.447, 81.475){\circle*{1.0}}
\put(20.447, 28.525){\circle*{1.0}}
\put(20.220, 82.030){\circle*{1.0}}
\put(20.220, 27.970){\circle*{1.0}}
\put(19.992, 82.586){\circle*{1.0}}
\put(19.992, 27.414){\circle*{1.0}}
\put(19.765, 83.141){\circle*{1.0}}
\put(19.765, 26.859){\circle*{1.0}}
\put(19.538, 83.697){\circle*{1.0}}
\put(19.538, 26.303){\circle*{1.0}}
\put(19.311, 84.252){\circle*{1.0}}
\put(19.311, 25.748){\circle*{1.0}}
\put(19.084, 84.807){\circle*{1.0}}
\put(19.084, 25.193){\circle*{1.0}}
\put(18.823, 85.347){\circle*{1.0}}
\put(18.823, 24.653){\circle*{1.0}}
\put(18.562, 85.888){\circle*{1.0}}
\put(18.562, 24.112){\circle*{1.0}}
\put(18.300, 86.428){\circle*{1.0}}
\put(18.300, 23.572){\circle*{1.0}}
\put(18.039, 86.968){\circle*{1.0}}
\put(18.039, 23.032){\circle*{1.0}}
\put(17.777, 87.508){\circle*{1.0}}
\put(17.777, 22.492){\circle*{1.0}}

\put(50,55){\vector(1,0){25}}
\put(50,62){\vector(4,1){25}} \put(50,48){\vector(4,-1){25}}
\multiput(50,84)(1.5,-1.5){4}{\vector(-1,-1){12}}
\put(40,5){\makebox(20,8)[l]{\bf (a)}}
\put(23.5,55){\oval(3,8){}}
\put(22,46){\makebox(0,8)[r]{$\small (\Delta S)_i$}}
\put(28,80){\makebox(0,8)[r]{$S$}}

\hspace{-3.5cm}
\put(228,80){\makebox(0,8)[r]{$S$}}
\put(222,46){\makebox(0,8)[r]{$\small (\Delta S)_i$}}
\put(223.5,55){\oval(3,8){}}
\put(245,5){\makebox(20,8)[l]{\bf (b)}}
\multiput(250,84)(1.5,-1.5){4}{\vector(-1,-1){12}}
\put(260,65){\vector(4,-1){25}} \put(260,45){\vector(4,1){25}}
\put(260,55){\vector(1,0){25}}

\put(225.000, 55.000){\circle*{1}}
\put(225.014, 55.600){\circle*{1.0}}
\put(225.014, 54.400){\circle*{1.0}}
\put(225.029, 56.200){\circle*{1.0}}
\put(225.029, 53.800){\circle*{1.0}}
\put(225.043, 56.799){\circle*{1.0}}
\put(225.043, 53.201){\circle*{1.0}}
\put(225.058, 57.399){\circle*{1.0}}
\put(225.058, 52.601){\circle*{1.0}}
\put(225.072, 57.999){\circle*{1.0}}
\put(225.072, 52.001){\circle*{1.0}}
\put(225.086, 58.599){\circle*{1.0}}
\put(225.086, 51.401){\circle*{1.0}}
\put(225.127, 59.198){\circle*{1.0}}
\put(225.127, 50.802){\circle*{1.0}}
\put(225.168, 59.796){\circle*{1.0}}
\put(225.168, 50.204){\circle*{1.0}}
\put(225.208, 60.395){\circle*{1.0}}
\put(225.208, 49.605){\circle*{1.0}}
\put(225.249, 60.993){\circle*{1.0}}
\put(225.249, 49.007){\circle*{1.0}}
\put(225.290, 61.592){\circle*{1.0}}
\put(225.290, 48.408){\circle*{1.0}}
\put(225.330, 62.191){\circle*{1.0}}
\put(225.330, 47.809){\circle*{1.0}}
\put(225.371, 62.789){\circle*{1.0}}
\put(225.371, 47.211){\circle*{1.0}}
\put(225.442, 63.385){\circle*{1.0}}
\put(225.442, 46.615){\circle*{1.0}}
\put(225.513, 63.981){\circle*{1.0}}
\put(225.513, 46.019){\circle*{1.0}}
\put(225.584, 64.577){\circle*{1.0}}
\put(225.584, 45.423){\circle*{1.0}}
\put(225.656, 65.172){\circle*{1.0}}
\put(225.656, 44.828){\circle*{1.0}}
\put(225.727, 65.768){\circle*{1.0}}
\put(225.727, 44.232){\circle*{1.0}}
\put(225.798, 66.364){\circle*{1.0}}
\put(225.798, 43.636){\circle*{1.0}}
\put(225.896, 66.956){\circle*{1.0}}
\put(225.896, 43.044){\circle*{1.0}}
\put(225.994, 67.548){\circle*{1.0}}
\put(225.994, 42.452){\circle*{1.0}}
\put(226.093, 68.140){\circle*{1.0}}
\put(226.093, 41.860){\circle*{1.0}}
\put(226.191, 68.731){\circle*{1.0}}
\put(226.191, 41.269){\circle*{1.0}}
\put(226.289, 69.323){\circle*{1.0}}
\put(226.289, 40.677){\circle*{1.0}}
\put(226.388, 69.915){\circle*{1.0}}
\put(226.388, 40.085){\circle*{1.0}}
\put(226.486, 70.507){\circle*{1.0}}
\put(226.486, 39.493){\circle*{1.0}}
\put(226.616, 71.093){\circle*{1.0}}
\put(226.616, 38.907){\circle*{1.0}}
\put(226.747, 71.678){\circle*{1.0}}
\put(226.747, 38.322){\circle*{1.0}}
\put(226.877, 72.264){\circle*{1.0}}
\put(226.877, 37.736){\circle*{1.0}}
\put(227.007, 72.850){\circle*{1.0}}
\put(227.007, 37.150){\circle*{1.0}}
\put(227.138, 73.436){\circle*{1.0}}
\put(227.138, 36.564){\circle*{1.0}}
\put(227.268, 74.021){\circle*{1.0}}
\put(227.268, 35.979){\circle*{1.0}}
\put(227.428, 74.600){\circle*{1.0}}
\put(227.428, 35.400){\circle*{1.0}}
\put(227.587, 75.178){\circle*{1.0}}
\put(227.587, 34.822){\circle*{1.0}}
\put(227.747, 75.756){\circle*{1.0}}
\put(227.747, 34.244){\circle*{1.0}}
\put(227.906, 76.335){\circle*{1.0}}
\put(227.906, 33.665){\circle*{1.0}}
\put(228.066, 76.913){\circle*{1.0}}
\put(228.066, 33.087){\circle*{1.0}}
\put(228.225, 77.492){\circle*{1.0}}
\put(228.225, 32.508){\circle*{1.0}}
\put(228.385, 78.070){\circle*{1.0}}
\put(228.385, 31.930){\circle*{1.0}}
\put(228.580, 78.637){\circle*{1.0}}
\put(228.580, 31.363){\circle*{1.0}}
\put(228.774, 79.205){\circle*{1.0}}
\put(228.774, 30.795){\circle*{1.0}}
\put(228.969, 79.773){\circle*{1.0}}
\put(228.969, 30.227){\circle*{1.0}}
\put(229.164, 80.340){\circle*{1.0}}
\put(229.164, 29.660){\circle*{1.0}}
\put(229.359, 80.908){\circle*{1.0}}
\put(229.359, 29.092){\circle*{1.0}}
\put(229.553, 81.475){\circle*{1.0}}
\put(229.553, 28.525){\circle*{1.0}}
\put(229.780, 82.030){\circle*{1.0}}
\put(229.780, 27.970){\circle*{1.0}}
\put(230.008, 82.586){\circle*{1.0}}
\put(230.008, 27.414){\circle*{1.0}}
\put(230.235, 83.141){\circle*{1.0}}
\put(230.235, 26.859){\circle*{1.0}}
\put(230.462, 83.697){\circle*{1.0}}
\put(230.462, 26.303){\circle*{1.0}}
\put(230.689, 84.252){\circle*{1.0}}
\put(230.689, 25.748){\circle*{1.0}}
\put(230.916, 84.807){\circle*{1.0}}
\put(230.916, 25.193){\circle*{1.0}}
\put(231.177, 85.347){\circle*{1.0}}
\put(231.177, 24.653){\circle*{1.0}}
\put(231.438, 85.888){\circle*{1.0}}
\put(231.438, 24.112){\circle*{1.0}}
\put(231.700, 86.428){\circle*{1.0}}
\put(231.700, 23.572){\circle*{1.0}}
\put(231.961, 86.968){\circle*{1.0}}
\put(231.961, 23.032){\circle*{1.0}}
\put(232.223, 87.508){\circle*{1.0}}
\put(232.223, 22.492){\circle*{1.0}}
\end{picture}

\vspace{-0.2cm}
\begin{center}
\begin{minipage}{10cm}
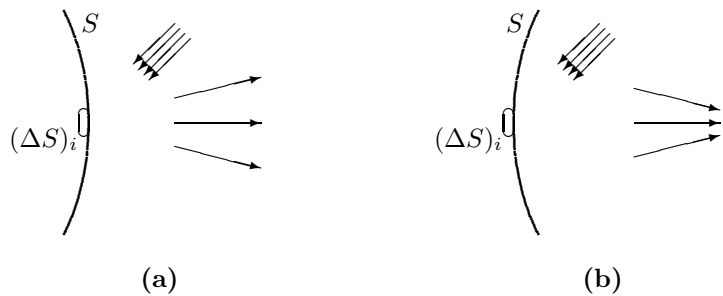
{Figure~3: Illustration of how boundaries `reflect' incident molecules. (a) If the surface is convex and rather smooth, the reflected molecules diverge. (b) The reflected molecules may converge.}
\end{minipage}
\end{center}

Are the diverging and converging molecules discussed in the last paragraph important to gas dynamics? Of course, they are. If the gas in question is a really rarefied one, almost all the molecules in it can be regarded as the ones produced directly by certain boundaries, which means that before a good equilibrium is established a great number of molecules need to be described by discontinuous distributions. For a more practical gas, we can at least say that the regions immediately near boundaries are very much filled with diverging and converging molecules abnormal to the Boltzmann equation, which is sufficient to challenge the standard methodologies in view of that the distributions in the inner region (assumed to exactly obey the Boltzmann equation for the discussion herein) has to be connected with the distributions caused by boundaries in one way 
or another.

We continue searching for cases in that distribution functions are not perfectly continuous. In Fig.~4a, on the
left side of the wall the space is filled with a dense gas and on the right side with a dilute gas (or just a
vacuum). It is obvious that the molecules leaking through a small hole on the wall will behave like diverging
molecules in the phase space. In Fig.~4b, a gas with a very high density is confined to a small spherical ball.
Suppose that the ball confinement suddenly vanishes at a moment (chosen as the initial time), and then the gas
expands almost freely. Obviously, these expanding molecules also behave like diverging molecules. Finally,
referring to Fig.~4c, consider two narrow beams of molecules meeting with each other in a small region. It can be
seen that the scattered molecules by the collisions there will diverge in the phase space. The examples presented
thus far tell us that molecule-to-boundary collisions, molecule-to-molecule collisions, boundary conditions and
initial conditions are all capable of making molecules diverge or converge in the phase space.

\hspace{0.09cm}
\setlength{\unitlength}{0.023in} 
\begin{picture}(100,70)
\put(32,41){\vector(1,0){18}}
\put(32,42.5){\vector(4,1){18}}
\put(32,39.5){\vector(4,-1){18}}

\multiput(18,24)(0,2){18}{\circle*{0.5}} 
\multiput(16,25)(0,2){17}{\circle*{0.5}} 
\multiput(14,24)(0,2){18}{\circle*{0.5}} 
\multiput(12,25)(0,2){17}{\circle*{0.5}} 
\multiput(10,24)(0,2){18}{\circle*{0.5}} 
\multiput(8,25)(0,2){17}{\circle*{0.5}} 
\multiput(20,20)(8,0){2}{\line(0,1){20}} 
\multiput(20,42)(8,0){2}{\line(0,1){20}} 
\multiput(20,40)(0,-1.8){10}{\line(1,0){8}} 
\multiput(20,42)(0,1.8){10}{\line(1,0){8}} 

\put(25,5){\makebox(0,8)[c]{\bf (a)}}

\hspace{-1.0cm}
\put(105,50){\vector(0,1){12}}
\put(105,32){\vector(0,-1){12}}
\put(114,41){\vector(1,0){12}}
\put(96,41){\vector(-1,0){12}}
\put(112,48){\vector(1,1){8}}
\put(98,34){\vector(-1,-1){8}}
\put(112,34){\vector(1,-1){8}}
\put(98,48){\vector(-1,1){8}}
\multiput(103,41)(1.2,0){5}{\circle*{0.5}} 
\multiput(103.5,42.5)(1.2,0){4}{\circle*{0.5}} 
\multiput(103.5,39.5)(1.2,0){4}{\circle*{0.5}} 
\put(108.000, 41.000){\circle*{1}}
\put(107.965, 41.398){\circle*{0.5}}
\put(107.534, 42.510){\circle*{0.5}}
\put(107.338, 42.858){\circle*{0.5}}
\put(106.405, 43.600){\circle*{0.5}}
\put(106.059, 43.800){\circle*{0.5}}
\put(104.878, 43.959){\circle*{0.5}}
\put(104.483, 43.898){\circle*{0.5}}
\put(103.374, 43.480){\circle*{0.5}}
\put(103.065, 43.226){\circle*{0.5}}
\put(102.327, 42.300){\circle*{0.5}}
\put(102.191, 41.923){\circle*{0.5}}
\put(102.033, 40.750){\circle*{0.5}}
\put(102.108, 40.357){\circle*{0.5}}
\put(102.581, 39.262){\circle*{0.5}}
\put(102.845, 38.961){\circle*{0.5}}
\put(103.811, 38.263){\circle*{0.5}}
\put(104.192, 38.139){\circle*{0.5}}
\put(105.379, 38.035){\circle*{0.5}}
\put(105.769, 38.123){\circle*{0.5}}
\put(106.842, 38.654){\circle*{0.5}}
\put(107.137, 38.924){\circle*{0.5}}
\put(107.791, 39.927){\circle*{0.5}}
\put(107.905, 40.310){\circle*{0.5}}

\put(105,5){\makebox(0,8)[c]{\bf (b)}}

\hspace{-1.0cm}
\put(164,48){\vector(3,-1){19}}
\put(164,34){\vector(3,1){19}}
\put(187,43){\vector(1,1){14}}
\put(186.2,43.5){\vector(3,4){12}}
\put(187.4,42.2){\vector(4,3){16}}
\put(186.9,39.3){\vector(1,-1){14}}
\put(186.4,38.5){\vector(3,-4){12}}
\put(187.4,39.8){\vector(4,-3){16}}
\put(185,5){\makebox(0,8)[c]{\bf (c)}}
\end{picture}

\vspace{-0.4cm}
\begin{center}
\begin{minipage}{10cm}
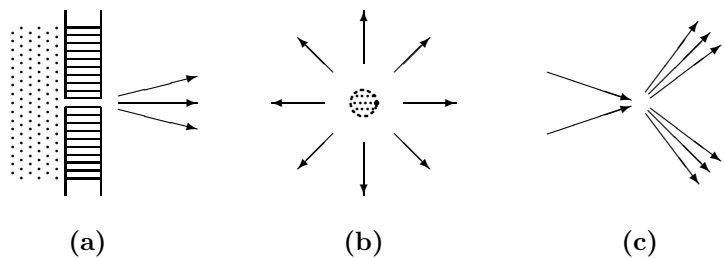
{Figure~4: (a) A gas leaking out of a small hole. (b) A gas expands freely after a certain time. (c) Two molecular beams collide in a small region.}
\end{minipage}
\end{center}

As far as the discontinuity issue is concerned, it is quite instructive to investigate the path invariance theorem
(\ref{invariance}) more thoroughly. Though the theorem is intimately related to the continuity assumption of
distribution function, it can, in a subtle and interesting way, lead us to betray the existing theory and admit
the necessity of using discontinuous distribution functions. To find out true implication of the path invariance,
we rewrite it as, with all external forces neglected for simplicity,
\begin{equation}  \label{source}
f(t,{\bf r},{\bf v})=f[t_0, {\bf r}_0,{\bf v}_0)] =f[t_0,{\bf r}-{\bf v}(t-t_0),
{\bf v}], \end{equation}
where $t_0$ is regarded as the initial time ($t_0<t$). Based on equation (\ref{source}), we may call ${\bf r}_0={\bf r}-{\bf v}(t-t_0)$ a source point and $\bf r$ an image point. We then assume the distribution function at $t_0$ to be known and try to determine the velocity distribution at a specific point ${\bf r}$ and at a later time $t$, denoted by $f_{t,\bf r}(v,\Omega)$ where $v=|{\bf v}|$ and the solid angle $\Omega$ specifies the direction of $\bf v$. From equation (\ref{source}), it is seen that for a certain value of $v$ all the source points of $f_{t,\bf r}(v,\Omega)$ distribute on the spherical shell that has the radius $v(t-t_0)$ and is centered at ${\bf r}$. Since $v$ is allowed to vary from zero to infinity, all points in the spatial space can serve as the source points of $f_{t,\bf r}(v,\Omega)$. In other words, the path invariance says that the velocity distribution at any spatial point is a condensed image of the entire spatial space; or, to say it in a converse way, any local spatial structure will constantly send its image to participate in forming the velocity distribution at every and each point elsewhere. A direct, and somewhat surprising, corollary of the above statements is that if a local structure is discontinuous no point in the space is truly immune from discontinuity.

\hspace{-0.5cm}
\setlength{\unitlength}{0.022in} 
\begin{picture}(100,85)

\multiput(52,18)(0,2){8}{\line(-1,0){25}} 
\multiput(52,18)(-25,0){2}{\line(0,1){14}} 

\multiput(54,18)(2,0){12}{\circle*{0.8}} 
\multiput(55,20)(2,0){12}{\circle*{0.8}} 
\multiput(54,22)(2,0){12}{\circle*{0.8}} 
\multiput(55,24)(2,0){12}{\circle*{0.8}} 
\multiput(54,26)(2,0){12}{\circle*{0.8}} 
\multiput(55,28)(2,0){12}{\circle*{0.8}} 
\multiput(54,30)(2,0){12}{\circle*{0.8}} 
\multiput(55,32)(2,0){12}{\circle*{0.8}}

\multiput(48,38.5)(20,25){2}{\line(0,1){13}}
\multiput(48,38.5)(20,25){2}{\line(1,0){4}}
\multiput(48,51.5)(20,25){2}{\line(-1,0){4}}
\multiput(48,45)(20,25){2}{\circle{18}} 
\multiput(148,45)(20,25){2}{\circle{18}} 

\put(153,25){\oval(44,15){}} 

\multiput(51.5,25)(-0.20,1.2){9}{\circle*{0.8}} 
\multiput(54.5,25)(-0.36,1.15){10}{\circle*{0.8}} 
\multiput(51.5,25)(0.445,1.2){30}{\circle*{0.8}} 
\multiput(54.5,25)(0.356,1.15){31}{\circle*{0.8}} 

\multiput(151.5,25)(-0.20,1.2){9}{\circle*{0.8}} 
\multiput(154.5,25)(-0.36,1.15){10}{\circle*{0.8}} 
\multiput(151.5,25)(0.445,1.2){30}{\circle*{0.8}} 
\multiput(154.5,25)(0.356,1.15){31}{\circle*{0.8}} 

\put(148.000, 45.000){\circle*{0.8}}
\put(148.165, 44.423){\circle*{0.8}}
\put(147.835, 45.577){\circle*{0.8}}
\put(148.267, 43.832){\circle*{0.8}}
\put(147.733, 46.168){\circle*{0.8}}
\put(148.357, 43.239){\circle*{0.8}}
\put(147.643, 46.761){\circle*{0.8}}
\put(148.448, 42.646){\circle*{0.8}}
\put(147.552, 47.354){\circle*{0.8}}
\put(148.549, 42.054){\circle*{0.8}}
\put(147.451, 47.946){\circle*{0.8}}
\put(148.665, 41.465){\circle*{0.8}}
\put(147.335, 48.535){\circle*{0.8}}
\put(148.822, 40.886){\circle*{0.8}}
\put(147.178, 49.114){\circle*{0.8}}
\put(149.033, 40.325){\circle*{0.8}}
\put(146.967, 49.675){\circle*{0.8}}
\put(149.348, 39.814){\circle*{0.8}}
\put(146.652, 50.186){\circle*{0.8}}
\put(149.821, 39.445){\circle*{0.8}}
\put(146.179, 50.555){\circle*{0.8}}
\put(150.381, 39.229){\circle*{0.8}}
\put(145.619, 50.771){\circle*{0.8}}
\put(150.973, 39.129){\circle*{0.8}}
\put(145.027, 50.871){\circle*{0.8}}
\put(151.570, 39.076){\circle*{0.8}}
\put(144.430, 50.924){\circle*{0.8}}
\put(152.169, 39.045){\circle*{0.8}}
\put(143.831, 50.955){\circle*{0.8}}

\put(168.000, 70.000){\circle*{0.8}}
\put(168.034, 69.401){\circle*{0.8}}
\put(167.966, 70.599){\circle*{0.8}}
\put(168.069, 68.802){\circle*{0.8}}
\put(167.931, 71.198){\circle*{0.8}}
\put(168.100, 68.203){\circle*{0.8}}
\put(167.900, 71.797){\circle*{0.8}}
\put(168.130, 67.604){\circle*{0.8}}
\put(167.870, 72.396){\circle*{0.8}}
\put(168.161, 67.004){\circle*{0.8}}
\put(167.839, 72.996){\circle*{0.8}}
\put(168.192, 66.405){\circle*{0.8}}
\put(167.808, 73.595){\circle*{0.8}}
\put(168.247, 65.808){\circle*{0.8}}
\put(167.753, 74.192){\circle*{0.8}}
\put(168.329, 65.213){\circle*{0.8}}
\put(167.671, 74.787){\circle*{0.8}}
\put(168.456, 64.627){\circle*{0.8}}
\put(167.544, 75.373){\circle*{0.8}}
\put(168.775, 64.119){\circle*{0.8}}
\put(167.225, 75.881){\circle*{0.8}}
\put(169.351, 63.948){\circle*{0.8}}
\put(166.649, 76.052){\circle*{0.8}}
\put(169.950, 63.920){\circle*{0.8}}
\put(166.050, 76.080){\circle*{0.8}}
\put(170.550, 63.921){\circle*{0.8}}
\put(165.450, 76.079){\circle*{0.8}}
\put(171.150, 63.929){\circle*{0.8}}
\put(164.850, 76.071){\circle*{0.8}}
\put(171.750, 63.936){\circle*{0.8}}
\put(164.250, 76.064){\circle*{0.8}}
\put(53,5){\makebox(0,8)[c]{\bf (a)}}
\put(153,5){\makebox(0,8)[c]{\bf (b)}}
\put(147,21){\makebox(0,8)[c]{${\bf r}_1$}}
\put(159,21){\makebox(0,8)[c]{${\bf r}_2$}}
\end{picture}

\begin{center}\vspace{-0.6cm}
\begin{minipage}{10cm}
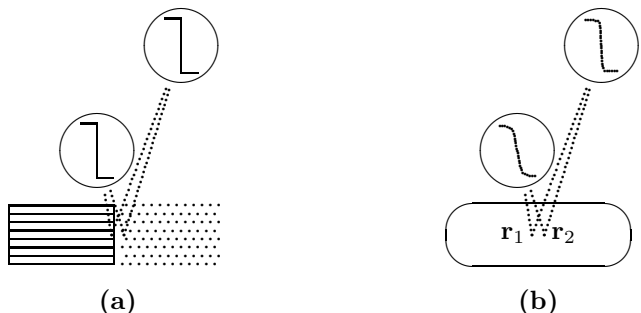
{Figure~5: At each point in space, the velocity distribution is formed as an image of structures elsewhere. The stuctures can be (a) a solid block surrounded by a gas; (b) a continuous gas with density gradient.}
\end{minipage}
\end{center}

Making things worse, even a continuous distribution can constantly send `bad images' to other places. Let's compare between Fig.~5a and Fig.5b. In Fig.~5a, there is a solid block surrounded by a continuous gas. The velocity distributions at other points, as images of the block and the gas, must involve corresponding discontinuity, as analyzed in the last paragraph. We then turn to Fig.~5b and assume that the distribution function in the oval region can be expressed by a local Maxwellian $n({\bf r})\exp(-mv^2/2T)$, where the temperature $T$ is constant and the symbol $n({\bf r})$ represents that the density there has a certain spatial gradient (but still perfectly contitinuous). From (\ref{source}), we find that the sharpness of the velocity distribution at an image point $\bf r$ outside the region can be characterized by
\begin{equation} \label{sharpness}
\left.\frac{\partial f({\bf v})}{\partial \theta}\right|_{\bf r}\approx
\frac{n({\bf r}_2)-n({\bf r}_1)}{\Delta \theta} e^{-mv^2/(2T)}, \end{equation}
where ${\bf r}_1$ and ${\bf r}_2$ are two source points in the oval region and $\theta$ is the angle formed by the two paths ${\bf r}-{\bf r}_1$ and ${\bf r}-{\bf r}_2$ (both are straight lines with no external force).
This expression informs us that as the image point $\bf r$ goes farther and farther away from the region (while holding ${\bf r}_1$ and ${\bf r}_2$ fixed), the profile of velocity distribution will have a sharper and sharper gradient in terms of the velocity angles and, finally, get almost the same shape as that caused discontinuous source structures. For convenience of discussion, we shall say that the velocity distributions illustrated in Fig.~5b are those involving quasi-discontinuity, though similar phenomena have been studied in mathematics under the title  `nonuniform continuity'\cite{analysis}.

The last two paragraphs manifest several interesting characteristics of gas dynamics. (i) Macroscopic structures constantly send their images (structure information) to shape and form microscopic structures elsewhere. (ii) When the structure information is conveyed through space it is condensed and sharpened progressively. (iii) Since a local distribution is a condensed image of its surrounding environment, it is improper and unnecessary to endow distribution functions with special continuity properties. (iv) Physical events make their  influences through one-dimensional trajectories, which is described by collisionless Newton's equations or Hamilton's equations.

We now wish to examine the discontinuous distribution function produced by a boundary of finite size in an analytical manner. Experimental facts tell us that collisions between molecules and boundaries cannot be truly elastic; stochastic and dissipative forces must get involved to a certain degree\cite{pathria}\cite{kogan}. To describe molecules `emitted' by such boundaries, it is adequate, in particular for a rarefied gas,  to define the local emission rate $\sigma$ in such a way that
\begin{equation}  \sigma dtdS dv d\Omega \end{equation}
represents the number of molecules that are reflected by the surface element $dS$ and emerge in the speed range $dv$ and in the solid angle range $d\Omega$ during the time interval $dt$. Referring Fig.~3 and dividing $S$ there into $N$ elements, we find that for the $i$th element of $S$ the reflected molecules are like ones emitted from a point source and the corresponding distribution function at a point ${\bf r}$ elsewhere can be expressed by, with external forces neglected,
\begin{equation}  \label{delta} f({\bf r},v,\Omega)=\frac{\sigma (\Delta S)_i} {|{\bf r}-{\bf r}_{0i}|^2 v^3 }V_i(v)\delta
(\Omega-\Omega_{{\bf r}-{\bf r}_{0i}}), \end{equation}
where $\Omega$ stands for a solid angle in the velocity space at ${\bf r}$, $V_i(v)$ is a function of $v$, ${\bf r}_{0i}$ is the position vector of $(\Delta S)_i$ and the solid angle $\Omega_{{\bf r}-{\bf r}_{0i}}$ takes the direction in that a reflected  molecule will continue to move from ${\bf r}$ after passing through its trajectory ${\bf r}_{0i}\rightarrow {\bf r}$. Since external forces are ignored, $\Omega_{{\bf r}-{\bf r}_{0i}}$ simply points in the direction of ${\bf r}-{\bf r}_{0i}$. In terms of ${\bf r}$ and $v$, this distribution function is continuous; in terms of $\Omega$, it is confined to a certain `point' $\Omega_{{\bf r}-{\bf r}_{0i}}$. In this sense, we wish to say that the distribution function (\ref{delta}) is a function of four dimensions.
The distribution function at ${\bf r}$ associated with all reflected molecules from all the area elements is
\begin{equation}  \label{delta1} f({\bf r},v,\Omega)=\sum\limits_{i}^{N}\frac{\sigma (\Delta S)_i} {|{\bf r}-{\bf r}_{0i}|^2 v^3 }V_i(v)\delta (\Omega-\Omega_{{\bf r}-{\bf r}_{0i}}). \end{equation}
Note that, on the usual understanding of distribution function, the summation symbol in (\ref{delta1}) is a formal one and we cannot replace it with the integral sign since each of $\Omega_{{\bf r}-{\bf r}_{0i}}$ represents a distinctive direction. Fig.~3 has shown that the molecules related to (\ref{delta1}) can diverge or converge in the phase space, mainly depending on how $S$ is arranged. As a limiting case, if $S$ is sufficiently small, equation (\ref{delta1}), being identical to (\ref{delta}), represents diverging molecules.

Though the above discussion appears to be trivial, many new concepts can arise from it. As one thing, allowing $\delta$-functions to express distribution functions shakes the prevailing belief that distribution functions must be continuous and gas dynamics must be analyzed by differential equations. As another, with the use of such $\delta$-functions, path information of molecules (which is  described by collisionless Newton's equations or Hamilton's equations) enters into the picture in a natural and treatable way. Finally, in order to relate this type of distribution function to the usual type of distribution function, we are actually compelled to invoke integral-type formulations.

Before finishing this subject, one issue may come into our mind. In fluid mechanics, several types of discontinuity, such as those related to shock waves, have been treated by finite difference schemes in a fairly successful way, why should we give so much concern to discontinuity in the kinetic regime? There are a number of things that can justify our attitude. Firstly, unlike what happens in fluid mechanics, discontinuities discussed in this paper spread out and involve almost all points in space.
Secondly, we have seen that the quasi-discontinuity produced by distribution functions will become sharper and sharper when propagating, and this type of dynamics cannot be dealt with by finite difference schemes.
Finally, all types of discontinuity presented in this section transfer along molecular paths, suggesting that path information must play a vital role in the would-be formalism, whereas path information is disregarded in almost all finite difference schemes of solving the Boltzmann equation (see the discussion in Sect.~ 2 also).

\section{The collision aspect of the Boltzmann equation}
The collisional operator of the Boltzmann equation seems to have been `clearly elaborated' in many textbooks. However, as will be seen, these elaborations actually involve vague descriptions and tacit assumptions. If we look at them with a mind less preoccupied, paradoxical concepts hidden in them will unveil themselves.

\hspace{0.2cm}
\setlength{\unitlength}{0.023in} 
\begin{picture}(100,58)

\put(28,44){\vector(2,-1){20}} 
\put(28,20){\vector(2,1){20}} 
\put(52,34){\vector(2,1){20}} 
\put(52,30){\vector(2,-1){20}} 

\multiput(49.5,33.5)(80,0){2}{\circle*{0.8}}
\multiput(51.5,33.5)(80,0){2}{\circle*{0.8}}
\multiput(50.5,33.3)(80,0){2}{\circle*{0.8}}
\multiput(49.5,30.5)(80,0){2}{\circle*{0.8}}
\multiput(51.5,30.5)(80,0){2}{\circle*{0.8}}
\multiput(50.5,30.7)(80,0){2}{\circle*{0.8}}
\put(50,9){\makebox(0,8)[c]{\bf (a)}}

\put(128,34){\vector(-2,1){20}} 
\put(128,30){\vector(-2,-1){20}} 
\put(152,44){\vector(-2,-1){20}} 
\put(152,20){\vector(-2,1){20}} 

\put(130,9){\makebox(0,8)[c]{\bf (b)}}

\put(40,27.5){\makebox(0,8)[c]{$\small {{\bf v}_2}$}}
\put(40,39){\makebox(0,8)[c]{$\small {\bf v}_1$}}
\put(60,27.5){\makebox(0,8)[c]{$\small {\bf v}_2^\prime$}}
\put(60,39){\makebox(0,8)[c]{$\small {\bf v}_1^\prime$}}

\put(118,27.5){\makebox(0,8)[c]{$\small {\bf -v}_2$}}
\put(118,39){\makebox(0,8)[c]{$\small {\bf -v}_1$}}
\put(138,27.5){\makebox(0,8)[c]{$\small {\bf -v}_2^\prime$}}
\put(138,39){\makebox(0,8)[c]{$\small {\bf -v}_1^\prime$}}
\end{picture}
\vskip-0.5cm
\begin{center}
\begin{minipage}{10cm}\vspace{-1.0cm}
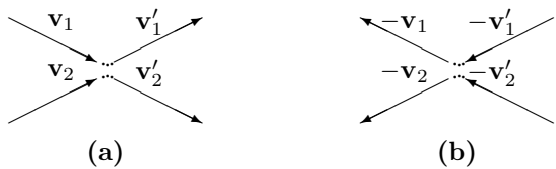
{Figure~6: Schematic of the time reversibility of two individual molecules. (a) The original collision; and (b) the inverse collision. }
\end{minipage}
\end{center}

To begin with, we recall the usual description of time reversal for a collision of two molecules with the same mass (still distinguishable according to classical mechanics). The initial and final velocities of the first molecule are denoted by ${\bf v}_1$ and ${\bf v}_1^\prime$ respectively; the initial and final velocities of the second molecule by ${\bf v}_2$ and ${\bf v}_2^\prime$ respectively. The following two collisions, referring to Fig.~6,
\begin{equation}  {\bf v}_1,{\bf v}_2 \rightarrow{\bf v}_1^\prime,{\bf v}_2^\prime \quad {\rm and}\quad
-{\bf v}_1^\prime,-{\bf v}_2^\prime \rightarrow -{\bf v}_1,-{\bf v}_2
\end{equation}
can be deemed as a pair of corresponding collisions. Under this understanding, the time reversibility of collision implies that if the first collision of the pair is physically possible then the second one must also be physically possible. This type of time reversibility has been solidly established in classical mechanics and we shall discuss it no more.

Now, we try to investigate whether or not there is time reversibility in terms of collisions between two molecular beams. Apparently, to the Boltzmann equation a time reversibility of this type is  much more relevant. In what follows, it will be shown that, much contrary to the conventional wisdom, no time reversibility of this type can be found out: neither an intuitive one, nor a mathematical one.

Intuitively, we may consider the following pair of corresponding pictures as a `candidate' for such time reversibility. The original collision picture is that two molecular beams with definite velocities collide with each other and the molecules scattered by collisions diverge in the spatial space with many different velocities, as shown in Fig.~7a. The inverse collision picture is that many different converging beams collide with each other and the molecules scattered by collisions form two molecular beams with two definite velocities, as shown in Fig.~7b. In no need of much discussion, we all know that the first picture makes sense in statistical mechanics, while the second one does not.

\hspace{-0.5cm}
\setlength{\unitlength}{0.023in} 
\begin{picture}(100,75)

\multiput(24,23)(-1,2){2}{\vector(2,1){22}} 
\multiput(24,52)(-1,-2){2}{\vector(2,-1){22}} 
\put(52,41){\vector(1,1){18}}
\put(53,40){\vector(3,2){20}}
\put(51,42){\vector(2,3){14}}

\put(52,34){\vector(1,-1){18}}
\put(53,35){\vector(3,-2){20}}
\put(51,33){\vector(2,-3){14}}
\put(50,2){\makebox(0,8)[c]{\bf (a)}}

\multiput(145,39.6)(1,2){2}{\vector(-2,1){22}}
\multiput(145,36.4)(1,-2){2}{\vector(-2,-1){22}}
\put(171,59.5){\vector(-1,-1){18}}
\put(174,53.5){\vector(-3,-2){20}}
\put(166,63.5){\vector(-2,-3){14}}

\put(171,16){\vector(-1,1){18}}
\put(174,22){\vector(-3,2){20}}
\put(166,12){\vector(-2,3){14}}
\put(150,2){\makebox(0,8)[c]{\bf (b)}}
\end{picture}
\vskip-0.3cm
\begin{center}
\begin{minipage}{10cm}\vspace{-0.2cm}
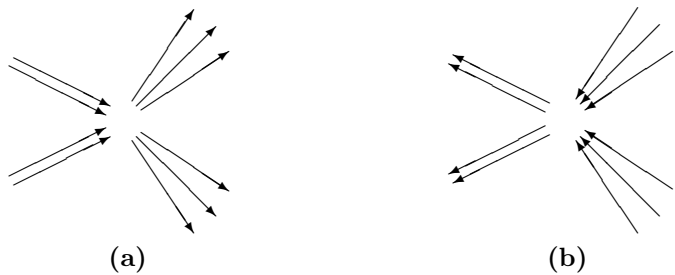
{Figure~7: A possible candidate of the time reversibility of beam-to-beam collision: (a) the original collisions; and (b) possible inverse collisions. }
\end{minipage}
\end{center}

Mathematically, according to the textbook treatment\cite{reif}, the time reversibility of beam-to-beam collision can be expressed by
\begin{equation} \label{equality} \sigma({\bf v}_1,{\bf v}_2 \rightarrow
{\bf v}_1^\prime, {\bf v}_2^\prime) = \sigma ({\bf v}_1^\prime,
{\bf v}_2^\prime \rightarrow {\bf v}_1, {\bf v}_2) ,\end{equation}
where the cross section $\sigma({\bf v}_1,{\bf v}_2 \rightarrow
{\bf v}_1^\prime, {\bf v}_2^\prime )$ is defined in such a way that, after collisions between a beam of type-1 molecules at ${\bf v}_1$ and a type-2 molecule at ${\bf v}_2$,
\begin{equation}  \label{sigma1}
N= \sigma({\bf v}_1,{\bf v}_2 \rightarrow {\bf v}_1^\prime,
{\bf v}_2^\prime ) d{\bf v}_1^\prime d{\bf v}_1^\prime
\end{equation}
represents the number of type-1 molecules emerging between ${\bf v}_1^\prime$ and ${\bf v}_1^\prime+d {\bf v}_1^\prime$ per unit incident flux and unit time, while the type-2 molecule emerges between ${\bf v}_2^\prime$ and ${\bf v}_2^\prime+d {\bf v}_2^\prime$; and
the cross section $\sigma({\bf v}_1^\prime ,{\bf v}_2^\prime  \rightarrow
{\bf v}_1, {\bf v}_2)$ is defined in the same way.

An unfortunate fact is that the concepts related to (\ref{equality}) and (\ref{sigma1}) suffer from fatal problems. For the collisions, the energy and momentum conservation laws state that
\begin{equation} \label{conservation}
{\bf v}_1^\prime+ {\bf v}_2^\prime ={\bf v}_1+{\bf v}_2 \equiv 2{\bf c}
\quad {\rm and} \quad |{\bf v}_2^\prime- {\bf v}_1^\prime| =
|{\bf v}_2-{\bf v}_1| \equiv u,
\end{equation}
where ${\bf c}$ is the velocity of the center-of-mass and $u$ is the relative speed of the two molecules. Fig.~8a shows how we can get ${\bf c}$ and $u$ from the initial velocities ${\bf v}_1$ and ${\bf v}_2$, while Fig.~8b shows how $\bf c$ and $u$ form constraints on the final velocities ${\bf v}_1^\prime$ and ${\bf v}_2^\prime$. It is clear that the final velocities of all scattered molecules must fall on a spherical shell $S$ of diameter $u=|{\bf u}|$ in the velocity space, which will be called the accessible shell. Keeping this accessible shell in mind, two misconcepts connected with the definition (\ref{sigma1}) manifest themselves. The first is that after $d{\bf v}_1^\prime$ is specified, specifying $d{\bf v}_2^\prime$ in the definition (\ref{sigma1}) is a work overdone. The second is that the cross section should not be defined in reference to a velocity volume element (like $d{\bf v}_1^\prime$ in the definition). If we insist on doing so, the resultant cross section will unpleasantly  vary from zero to infinity, depending on the shape and size of the employed velocity volume element.

\hspace{5.6cm}
\setlength{\unitlength}{0.016in}
\begin{picture}(200,143)

\put(-120,85){\vector(3,-1){84.5}} 
\put(-120,85){\vector(1,1){27.5}}
\put(-120,85){\vector(1,0){56.5}}
\multiput(-91.5,113.5)(1.5,-1.5){36}{\circle*{1.2}}
\put(-86.5,113.5){\vector(1,-1){56.5}}
\put(-86.5,113.5){\circle*{3}}
\put(-106,109){\makebox(35,8)[l]{${\bf v}_1$}}
\put(-58,53){\makebox(35,8)[l]{${\bf v}_2$}}
\put(-90,77){\makebox(35,8)[c]{${\bf c}$}}
\put(-68,80){\makebox(35,8)[c]{${\bf u}$}}

\put(-80,30){\makebox(0,8)[c]{\bf (a)}}

\hspace{-2cm}
\put(98,30){\makebox(0,8)[c]{\bf (b)}}

\put(45.5,85){\vector(1,0){56.5}}
\multiput(103,85)(-1.5,-1.5){19}{\circle*{1.2}}
\multiput(103,85)(1.5,1.5){19}{\circle*{1.2}}

\put(45.5,85){\vector(3,1){84.5}} \put(45.5,85){\line(5,1){92}} \put(45.5,85){\vector(2,1){72.5}}
\put(136.5,103.4){\vector(4,1){1}} \put(45.5,85){\vector(4,3){52.5}} \put(45.5,85){\vector(1,-1){28.5}}
\put(45.5,85){\vector(1,1){28.5}}

\put(56,111){\makebox(35,8)[l]{$({\bf v}_1^\prime)$}}
\put(88,127){\makebox(35,8)[l]{$({\bf v}_1^\prime)$}}
\put(113,124){\makebox(35,8)[l]{$({\bf v}_1^\prime)$}}
\put(132,114){\makebox(35,8)[l]{${\bf v}_1^\prime$}}
\put(139,102){\makebox(35,8)[l]{$({\bf v}_1^\prime)$}}
\put(61,51){\makebox(35,8)[l]{${\bf v}_2^\prime$}}
\put(69,77){\makebox(35,8)[c]{${\bf c}$}}
\put(75,65){\makebox(35,8)[c]{$u$}}
\put(133,50){\makebox(35,8)[l]{$S$}}

\put(143.00, 85.00){\circle *{1.2}}
\put(142.78, 89.18){\circle *{1.2}}
\put(142.13, 93.32){\circle *{1.2}}
\put(141.04, 97.36){\circle *{1.2}}
\put(139.54, 101.27){\circle *{1.2}}
\put(137.64, 105.00){\circle *{1.2}}
\put(135.36, 108.51){\circle *{1.2}}
\put(132.73, 111.77){\circle *{1.2}}
\put(129.77, 114.73){\circle *{1.2}}
\put(126.51, 117.36){\circle *{1.2}}
\put(123.00, 119.64){\circle *{1.2}}
\put(119.27, 121.54){\circle *{1.2}}
\put(115.36, 123.04){\circle *{1.2}}
\put(111.32, 124.13){\circle *{1.2}}
\put(107.18, 124.78){\circle *{1.2}}
\put(103.00, 125.00){\circle *{1.2}}
\put(98.82, 124.78){\circle *{1.2}}
\put(94.68, 124.13){\circle *{1.2}}
\put(90.64, 123.04){\circle *{1.2}}
\put(86.73, 121.54){\circle *{1.2}}
\put(83.00, 119.64){\circle *{1.2}}
\put(79.49, 117.36){\circle *{1.2}}
\put(76.23, 114.73){\circle *{1.2}}
\put(73.27, 111.77){\circle *{1.2}}
\put(70.64, 108.51){\circle *{1.2}}
\put(68.36, 105.00){\circle *{1.2}}
\put(66.46, 101.27){\circle *{1.2}}
\put(64.96, 97.36){\circle *{1.2}}
\put(63.87, 93.32){\circle *{1.2}}
\put(63.22, 89.18){\circle *{1.2}}
\put(63.00, 85.00){\circle *{1.2}}
\put(63.22, 80.82){\circle *{1.2}}
\put(63.87, 76.68){\circle *{1.2}}
\put(64.96, 72.64){\circle *{1.2}}
\put(66.46, 68.73){\circle *{1.2}}
\put(68.36, 65.00){\circle *{1.2}}
\put(70.64, 61.49){\circle *{1.2}}
\put(73.27, 58.23){\circle *{1.2}}
\put(76.23, 55.27){\circle *{1.2}}
\put(79.49, 52.64){\circle *{1.2}}
\put(83.00, 50.36){\circle *{1.2}}
\put(86.73, 48.46){\circle *{1.2}}
\put(90.64, 46.96){\circle *{1.2}}
\put(94.68, 45.87){\circle *{1.2}}
\put(98.82, 45.22){\circle *{1.2}}
\put(103.00, 45.00){\circle *{1.2}}
\put(107.18, 45.22){\circle *{1.2}}
\put(111.32, 45.87){\circle *{1.2}}
\put(115.36, 46.96){\circle *{1.2}}
\put(119.27, 48.46){\circle *{1.2}}
\put(123.00, 50.36){\circle *{1.2}}
\put(126.51, 52.64){\circle *{1.2}}
\put(129.77, 55.27){\circle *{1.2}}
\put(132.73, 58.23){\circle *{1.2}}
\put(135.36, 61.49){\circle *{1.2}}
\put(137.64, 65.00){\circle *{1.2}}
\put(139.54, 68.73){\circle *{1.2}}
\put(141.04, 72.64){\circle *{1.2}}
\put(142.13, 76.68){\circle *{1.2}}
\put(142.78, 80.82){\circle *{1.2}}
\end{picture}
\vskip -1.5cm
\begin{center}
\begin{minipage}{10cm}\vspace{-1.4cm}
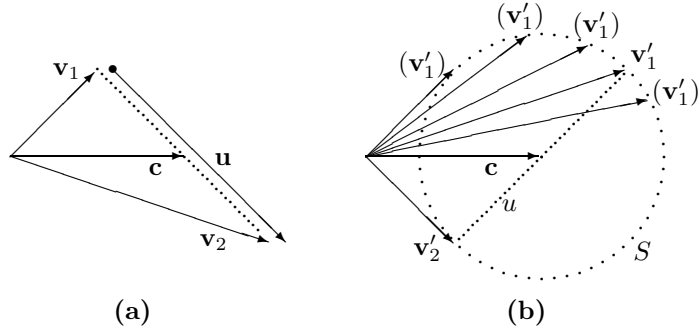
{Figure~8: Constraints imposed by the energy and momentum conservation
 laws. (a) The velocity
 ${\bf c}$ and $u=|{\bf u}|$ are determined by the initial velocities.
(b) The final velocities of the scattered molecules ${\bf v}_1^\prime$
  and ${\bf v}_2^\prime$ must fall on the accessible shell $S$ of diameter
$u$. }
\end{minipage}
\end{center}

\hspace{0.0cm}
\setlength{\unitlength}{0.014in} 
\begin{picture}(300,138)

\multiput(85,86)(90,0){3}{\makebox(8,8)[c]{$S$}}
\put(57,35){\makebox(8,8)[c]{\bf (a)}}
\put(147,35){\makebox(8,8)[c]{\bf (b)}}
\put(237,35){\makebox(8,8)[c]{\bf (c)}}

\put(46,105){\makebox(30,8)[c]{$d{\bf v}_1^\prime$}}
\put(136,118){\makebox(30,8)[c]{$d{\bf v}_1^\prime$}}
\put(226,105){\makebox(30,8)[c]{$d{\bf v}_1^\prime$}}

\put(150,95){\oval(10,40){}}
\put(240,95){\oval(40,10){}}
\put(60,95){\circle{15}}
\multiput(95.00,60.00)(90,0){3}{\circle*{1}}
\multiput(95.00,60.00)(90,0){3}{\circle*{1}}
\multiput(94.89,62.82)(90,0){3}{\circle*{1}}
\multiput(94.55,65.61)(90,0){3}{\circle*{1}}
\multiput(93.98,68.38)(90,0){3}{\circle*{1}}
\multiput(93.20,71.08)(90,0){3}{\circle*{1}}
\multiput(92.20,73.72)(90,0){3}{\circle*{1}}
\multiput(90.99,76.27)(90,0){3}{\circle*{1}}
\multiput(89.58,78.71)(90,0){3}{\circle*{1}}
\multiput(87.98,81.03)(90,0){3}{\circle*{1}}
\multiput(86.20,83.21)(90,0){3}{\circle*{1}}
\multiput(84.25,85.24)(90,0){3}{\circle*{1}}
\multiput(82.14,87.11)(90,0){3}{\circle*{1}}
\multiput(79.88,88.80)(90,0){3}{\circle*{1}}
\multiput(77.50,90.31)(90,0){3}{\circle*{1}}
\multiput(75.00,91.62)(90,0){3}{\circle*{1}}
\multiput(72.41,92.73)(90,0){3}{\circle*{1}}
\multiput(69.74,93.62)(90,0){3}{\circle*{1}}
\multiput(67.00,94.29)(90,0){3}{\circle*{1}}
\multiput(64.22,94.74)(90,0){3}{\circle*{1}}
\multiput(61.41,94.97)(90,0){3}{\circle*{1}}
\multiput(58.59,94.97)(90,0){3}{\circle*{1}}
\multiput(55.78,94.74)(90,0){3}{\circle*{1}}
\multiput(53.00,94.29)(90,0){3}{\circle*{1}}
\multiput(50.26,93.62)(90,0){3}{\circle*{1}}
\multiput(47.59,92.73)(90,0){3}{\circle*{1}}
\multiput(45.00,91.62)(90,0){3}{\circle*{1}}
\multiput(42.50,90.31)(90,0){3}{\circle*{1}}
\multiput(40.12,88.80)(90,0){3}{\circle*{1}}
\multiput(37.86,87.11)(90,0){3}{\circle*{1}}
\multiput(35.75,85.24)(90,0){3}{\circle*{1}}
\multiput(33.80,83.21)(90,0){3}{\circle*{1}}
\multiput(32.02,81.03)(90,0){3}{\circle*{1}}
\multiput(30.42,78.71)(90,0){3}{\circle*{1}}
\multiput(29.01,76.27)(90,0){3}{\circle*{1}}
\multiput(27.80,73.72)(90,0){3}{\circle*{1}}
\multiput(26.80,71.08)(90,0){3}{\circle*{1}}
\multiput(26.02,68.38)(90,0){3}{\circle*{1}}
\multiput(25.45,65.61)(90,0){3}{\circle*{1}}
\multiput(25.11,62.82)(90,0){3}{\circle*{1}}
\end{picture}
\vskip-1.5cm
\begin{center}
\begin{minipage}{10cm}\vspace{-1.6cm}
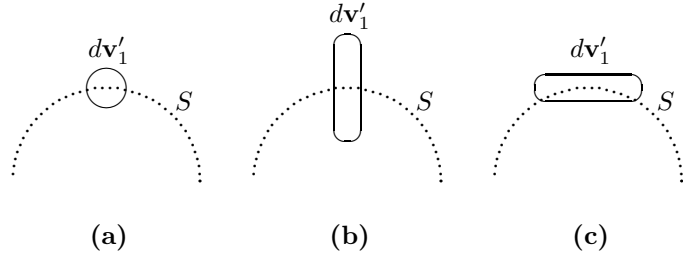
{Figure~9: Possible velocity volume elements $d{\bf v}_1^\prime$ are given, each of which encloses a small part of the accessible surface $S$. They can be (a) a spherical ball, (b) a slim cylinder, and (c) a tabular cylinder.}
\end{minipage}
\end{center}

To see the conclusion given in the last paragraph more vividly, look at Fig.~9, in which several possible shapes have been given to $d{\bf v}_1^\prime$. For the situation shown in Fig.~9a, the number of molecules entering the  spherical ball, chosen to be $d{\bf v}_1^\prime$, per unit time can be expressed by $N\approx \rho \pi r^2$, where $\rho$ is the area
density of molecules on the accessible shell caused by unit flux of type-1 molecules and $r$ is the radius of the spherical ball $d{\bf v}_1^\prime$. By noting that $\rho$ must be finite, we find that the cross section defined by (\ref{sigma1}) is equal to, with $d{\bf v}_2^\prime$ neglected,
\begin{equation}
\sigma=\frac N{d{\bf v}_1^\prime}=\frac{\rho \pi r^2} {4\pi r^3/3}= \frac{3\rho}{4r},
\end{equation}
and it tends to infinity as $r\rightarrow 0$. Nevertheless, if the volume element $d{\bf v}_1^\prime$ is chosen to be the one in Fig.~9b, then $\sigma$ tends to zero; if the volume element $d{\bf v}_1^\prime$ is chosen to be the one in Fig.~9c, $\sigma$ tends to infinity again. These examples simply mean that the cross section defined by (\ref{sigma1}) is, in fact, an ill-defined quantity, and thus the time reversibility (\ref{equality}) defined by such cross sections is also groundless.

It is now clear that the Boltzmann collisional operator indeed involves misconcepts. An interesting and essential question arises: Can we still formulate collisional effects in the context of the standard theory? In the rest of this section we try to get an answer to this question by reinvestigating collisional effects in the three-dimensional velocity space and in the 
six-dimensional phase space. Surprisingly enough, such investigations will give us nothing but more paradoxes.

Let's first investigate the number of the molecules that enter a volume element in 
a velocity space, denoted by $d{\bf v}_1$ herein. This type of investigation is not truly relevant, since the Boltzmann equation is supposed to be  cast in the full six-dimensional phase space, but we take this approach anyway, in order to see where we arrive at.
Following the usual methodology, we consider two molecular beams with velocities ${\bf v}_1^\prime$ and ${\bf v}_2^\prime$ (called type-1 and type-2 molecules respectively), and examine how many type-1 molecules emerge, after collisions, in the range between ${\bf v}_1$ and ${\bf v}_1+d{\bf v}_1$. As the discussion after (\ref{conservation}) has already manifested, because of the energy and momentum conservation laws, all the type-1 molecules scattered by collisions must fall 
on the accessible shell. If we insist on knowing the density of the molecules entering the volume element $d{\bf v}_1$, we get a result that varies drastically from zero to infinity, depending on the position, size and shape of $d{\bf v}_1$. 

We now turn our attention from the velocity space to the phase space and try to examine the net change of the molecular number in a six-dimensional phase volume element due to collisions. For the reason that has been clarified, the following  examination will be done without getting help from the cross section $\sigma({\bf v}_1,{\bf v}_2 \rightarrow {\bf v}_1^\prime, {\bf v}_2^\prime )$.

Firstly,  the molecules leaving $d{\bf r} d{\bf v}_1$ during $dt$ due to collisions are of our concern. Rather conventionally, we identify $f({\bf v}_1)d{\bf v}_1$ and $f({\bf v}_2)d{\bf v}_2$ as two molecular beams. By switching to the 
center-of-mass frame and studying the collisions related to the two beams, the number of the collisions in $d{\bf r}$ during $dt$ can be directly obtained as
\begin{equation}  dtd{\bf r}\int_\Omega d\Omega u\sigma(\Omega) [f({\bf v}_1)d{\bf v}_1]
 [f({\bf v}_2)d{\bf v}_2], \end{equation}
where $\sigma(\Omega)$ is well-defined in the center-of-mass frame\cite{landau}. Integrating the above formula over ${\bf v}_2$ yields the total number of the collisions taking place in
$d{\bf r} d{\bf v}_1$ during $dt$
\begin{equation}  \label{collisionn}
dtd{\bf r} d{\bf v}_1\int_{\Omega,{\bf v}_2} u f({\bf v}_1)
f({\bf v}_2)\sigma(\Omega) d\Omega d{\bf v}_2 .
\end{equation}
Until now, every step is almost the same as that in the standard approach,
and the comparison between this expression and the Boltzmann collisional operator seems to state that things go and will go as smoothly as expected. However, a careful inspection will tell us different stories. The form of $dt d{\bf r} d{\bf v}_1$ in expression (\ref{collisionn}) reminds us that $dt$, $d{\bf r}$ and $d{\bf v}_1$ are three differentials; in obtaining the Boltzmann equation, each of them must be allowed to approach zero independently. Conflicting conclusions will be drawn when different ways of taking these limits are adopted. One of the conclusions is that: if
$|d{\bf r}|>>|{\bf v}_1dt|$, where $|d{\bf r}|$ stands for the length scale of $d{\bf r}$, almost all the molecules not suffering collisions will stay inside $d{\bf r} d{\bf v}_1$ during $dt$ and the molecules expressed by (\ref{collisionn}) are indeed the ones that get velocity changes due to collisions and leave $d{\bf r} d{\bf v}_1$ during $dt$. But, another conclusion is quite opposite: if $|d{\bf r}|<<|{\bf v}_1dt|$, all the molecules initially inside $d{\bf r} d{\bf v}_1$  will get out of $d{\bf r} d{\bf v}_1$ in a time much shorter than $dt$, irrespective of suffering collisions or not (see Fig.~10a).
In the latter case, it is meaningless to say that the molecules described by  
(\ref{collisionn}) are those leaving the volume element during $dt$ just due to collisions. In other words, to make (\ref{collisionn}) relevant, we have no choice but to assume that $|d{\bf r}|>>|{\bf v}_1dt|$. An unfortunate fact is that no sound reason can be found out for that we can prefer this assumption to its converse.

\hspace{-2.0cm}
\setlength{\unitlength}{0.028in} 
\begin{picture}(80,80)
\put(57,57){\vector(1,1){15}} 
\multiput(56.5,56.5)(-1,-1){6}{\circle*{0.5}} 

\multiput(53.5,43.5)(0.8,-1.2){13}{\circle*{0.5}} 
\multiput(55,45)(1,-1){13}{\circle*{0.5}} 
\multiput(56,46.5)(1.2,-0.8){13}{\circle*{0.5}} 
\put(39,59){\makebox(20,8)[c]{$d{\bf r}d{\bf v}_1$}}
\put(61,60){\makebox(20,8)[c]{${\bf v}_1dt$}}
\put(63,25){\makebox(20,8)[c]{$({\bf v}_1^\prime)$}}

\put(70,36.6){\vector(3,-2){1}} 
\put(66.5,32.5){\vector(1,-1){1}} 
\put(62.5,29){\vector(2,-3){1}} 
\put(50,50){\circle{20}}
\put(57,16){\makebox(0,8)[c]{\bf (a)}}

\put(124,59){\makebox(20,8)[c]{$d{\bf r}d{\bf v}_1$}}
\put(156,60){\makebox(0,8)[c]{${\bf v}_1dt$}}
\put(117.5,42){\makebox(0,8)[c]{$({\bf v}_1^\prime)$}}
\put(152,39){\makebox(0,8)[c]{$({\bf v}_2)$}}
\put(130,30){\makebox(0,8)[c]{$({\bf v}_2^\prime)$}}
\put(135,50){\circle{20}}

\put(135,16){\makebox(0,8)[c]{\bf (b)}}
\put(121,50){\vector(1,0){0.5}}
\put(142,57){\vector(1,1){15}} 
\multiput(141.5,56.5)(-1,-1){6}{\circle*{0.5}} 
\put(134.0,50.3){\circle*{0.5}}
\put(135.4,50.7){\circle*{0.5}}
\multiput(132.6,50)(-1.4,0){19}{\circle*{0.5}} 

\multiput(135,47)(-1,-1){20}{\circle*{0.5}} 
\put(125,37){\vector(1,1){1}}
\put(136.35,47.3){\circle*{0.5}}
\put(137.7,47.5){\circle*{0.5}}
\multiput(139.1,47.6)(1.41,0){19}{\circle*{0.5}} 
\put(150,47.6){\vector(1,0){1}}
\end{picture}

\begin{center}
\vskip-1.0cm
\begin{minipage}{10cm}
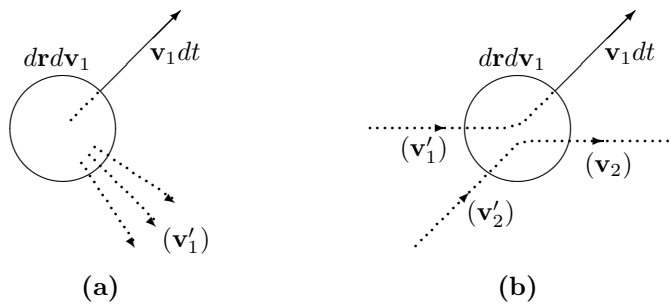
{Figure~10: Schematic of what happens in an infinitesimal element $d{\bf r}d{\bf v}_1$ during $dt$.
(a) If $|d{\bf r}|<<|{\bf v}_1dt|$ all molecules will leave $d{\bf r}$ almost instantly. (b) If $|d{\bf r}|<<|{\bf v}_1dt|$ most of molecules produced by collisions will not stay within $d{\bf r}$ after $dt$.}
\end{minipage}
\end{center}

Then, we examine molecules entering a phase volume element $d{\bf r} d{\bf v}_1$ during $dt$ due to collisions. Following the textbook treatment, two beams with velocities ${\bf v}_1^\prime$ and ${\bf v}_2^\prime$ are assumed to collide within a spatial volume element $d{\bf r}$ during $dt$, giving contribution to the molecules expressed by $f({\bf r},{\bf v}_1,t) d {\bf r} d{\bf v}_1$. Since no time reversibility can be utilized as a `shortcut', we have to try the task from scratch. It turns out that not one but many paradoxes will be in front of us. As the first paradox, if we assume, by adopting the customary thought, that only collisions taking place within $d{\bf r}$ are worth considering, then the size and shape of $d{\bf r}$ will become irrelevant and the six-dimensional collision issue will be reduced to the three-dimensional collision issue in the velocity space. As have already been discussed, the number of the molecules entering a 
three-dimensional velocity volume element is an ill-defined quantity and we have no way to determine it definitely. As the second paradox, even if we were able to formulate the relevant molecular number produced by collisions within $d{\bf r}d{\bf v}_1$ during $dt$, we would still encounter serious difficulties. For instance, we again need to let $dt$, $d{\bf r}$ and $d{\bf v}_1$ approach zero. If $|d{\bf r}|<<|{\bf v}_1 dt|$, all produced molecules will `instantly' leave $d{\bf r}d{\bf v}_1$ (see Fig.~10b) and cannot be treated as a true contribution to $f({\bf r},{\bf v}_1,t)d{\bf r}d{\bf v}_1$. As the third paradox, molecules produced by local collisions must diverge in the phase space, as revealed in the last section; such molecules have to be treated with care and cannot be regarded as a regular contribution to $f({\bf r},{\bf v}_1,t) d {\bf r} d{\bf v}_1$.

The discussions in this section, as well as some of those in the last section, suggest one essential and delicate thing: the phase space is a space much more complex than it appears and the collective behavior of molecules in it has to be studied in a careful and comprehensive way. The phase space, in which $x,y,z$ and $p_x, p_y, p_z$ are regarded as variables independent of each other, has been employed for many years and has served as a good mental and mathematical tool in classical mechanics. Its success is actually related to the existence of Hamilton's equations. With help of Hamilton's equations, the true correlation between $t,{\bf r}$ and ${\bf v}$ will be restored when the solution of Hamilton's equations is obtained. In statistical mechanics, the situation is subtly different. As has been illustrated in this section, to formulate collisional dynamics, it is sometimes necessary to consider the correlation between $t,{\bf r}$ and ${\bf v}$ and sometimes not necessary to do so. Before and after collisions, the molecules obey Hamilton's equations for individual molecules, in which $dt$, $d{\bf r}$ and $d{\bf v}$ indeed correlate. When collisions take place, the molecules are governed by laws of beam-to-beam collisions, in which certain types of statistical assumptions have to be adopted and Hamilton's equations for individual molecules do not play an essential role. Though the standard collisional operator tries hard to combine these two mechanisms, some conflicts remain there.

\section{Conclusions}
In summary, two paradoxical aspects of the standard Boltzmann equation have been presented. The first paradoxical aspect is related to the usual concept of distribution function. An tacit assumption of the existing kinetic equations is that distribution functions, though describing discrete molecules, must be mentally and practically continuous. The discussions in this paper, however, show that distribution functions of realistic gases can have complex structures over infinitesimal range, and such distribution functions do not keep invariant along a molecule's path even if the local collisions can be ignored. The second paradoxical aspect is related to the standard collisional operator. It is pointed out that the time reversibility in terms of beam-to-beam collisions is actually fictitious. Along this line, it is unveiled that there are unsurmountable difficulties in terms of formulating molecules entering and leaving an infinitesimal phase volume element. All these things pinpoint one thing in common: it is not possible to formulate collective behavior of molecules in the phase space mainly based on a differential-type analysis.

Many more fundamental questions arise sharply. Some of our recent works
try to make more analyses and put forward alternative approaches\cite{chen1,
chen2}. With help
of a development in quantum mechanics\cite{chen3}, part of the discussion in
this paper can be extended to the regime of quantum statistical physics.

This paper is supported by School of Science,
BUAA, PRC and by Education Ministry, PRC.


\begin{thebibliography}{99}
\bibitem{pathria} See, for instance, the
 preface by R. K. Pathria, {\it Statistical Mechanics}, (Pergamon Press,
1972), where a mention is made of the historical debate concerning the
Boltzmann equation.
\end{thebibliography}
\end{document}